\documentclass[aps,prb,twocolumn,groupedaddress,superscriptaddress,showpacs]{revtex4}
\usepackage{graphicx}

\begin{document}

\title{Bilayer graphene dual-gate nanodevice: An {\it ab initio} simulation}

\author{J. E. Padilha}
\email[]{padilha@if.usp.br}
\affiliation{Instituto de F\'isica, Universidade de S\~ao Paulo, CP 66318, 05315-970, S\~ao Paulo, SP, Brazil.}

\author{Matheus P. Lima}
\email[]{mplima@if.usp.br}
\affiliation{Instituto de F\'isica, Universidade de S\~ao Paulo, CP 66318, 05315-970, S\~ao Paulo, SP, Brazil.}

\author{Ant\^onio J. R. da Silva}
\email[]{ajrsilva@if.usp.br}
\affiliation{Instituto de F\'isica, Universidade de S\~ao Paulo, CP 66318, 05315-970, S\~ao Paulo, SP, Brazil.}
\affiliation{Laboratorio Nacional de Luz S\'incrotron - LNLS, CP 6192, 13083-970, Campinas, SP, Brazil.}

\author{A. Fazzio}
\email[]{fazzio@if.usp.br}
\affiliation{Instituto de F\'isica, Universidade de S\~ao Paulo, CP 66318, 05315-970, S\~ao Paulo, SP, Brazil.}

\date{\today}

\begin{abstract}

We study the electronic transport properties of a dual-gated bilayer
graphene nanodevice via first principles calculations. We
investigate the electric current as a function of gate length and
temperature. Under the action of an external electrical field we
show that even for gate lengths up $100$~\AA, a non zero current is
exhibited. The results can be explained by the presence of a
tunneling regime due the remanescent states in the gap. We also
discuss the conditions to reach the charge neutrality point in a
system free of defects and extrinsic carrier doping.

\end{abstract}


\maketitle

There is a consensus in the scientific community that the scaling
down of silicon-based metal-oxide-semiconductor field-effect transistors (MOSFETs) is approaching its limits. Therefore,
there is a pursuit for different materials to replace the silicon
paradigm\cite{gr-trans}. Among the potential candidates, graphene
has attracted the attention of research groups after the seminal
work of the Manchester group\cite{graph-first}. Graphene is a truly two-dimensional
(2D) material, with a honeycomb structure, and is a zero band-gap
semiconductor. The valence and conduction bands close to the Fermi
level are cone shaped, with a linear energy-momentum relation at the
two $k$ points ($K$ and $K^\prime$) of the Brillouin zone, called
Dirac points\cite{castro_RMP}. As a result, graphene has several
properties that make it very interesting to use in the manufacture of devices, such as its very
high electronic mobilities, up to $200~000~cm^{2}V^{-1}s^{-1}$
\cite{mobility1,mobility2}, with electrons and holes behaving as
massless fermions near the $K$ and $K^\prime$ points. Thus, charge
carriers can travel for micrometers at room temperature without any
scattering.

However, a drawback in using graphene as an electronic material, such
as in logic devices, is its lack of an energy gap. A number of
different approaches have been proposed to open such a bandgap on
graphene\cite{gap-gr1,gap-gr2,gap-gr3,gap-gr4,gap-gr5}. Another way
to create the gap is to use bilayer graphene, breaking the inversion
symmetry through an external perturbation, such as the application
of an electrical field perpendicular to the layers. Moreover, this
band gap can be tuned by varying the field strength. This fact has
been confirmed by both photoemission and optical absorption
experiments\cite{tunable1,tunable2,tunable3}. A few theoretical
studies have also investigated this approach to open a band gap in
bilayer graphene via the application of an electric field along the whole system\cite{teo-gap-bl1,teo-gap-bl2}. 
Recently, Avouris' group observed a transport band gap in biased bilayer
graphene\cite{on-off}, also demonstrating its potential for
applications in digital electronics. Thus, understanding the
intrinsic properties of devices based on bilayer graphene in
nano-dimensions is important for future developments in the area.

In this Brief Report we present investigations of the electronic transport
properties of a dual-gated bilayer graphene nanodevice, as a
function of the gate length ($L_{gate}$) and temperature, via first-principles 
calculations. We show an exponential decrease of the
current as a function of the gate length, which is characteristic of a
tunneling regime. As a result, the system does not reach a zero
current. Also, we discuss the conditions to achieve the charge
neutrality point (CNP) in a system that is not affected by defects
and extrinsic carrier doping. We present results for electric fields
comparable to the experimental electric displacement fields,
$1.0V/nm<|E|< 5.0V/nm$, for room temperature and $4.5K$. We use the
Landauer-B\"uttiker model, with a Hamiltonian generated through {\it
ab initio} density functional theory (DFT),\cite{dft1} coupled
with a non equilibrium Green's function formalism in a fully
self-consistent procedure (NEGF-DFT)\cite{negf1}. To
investigate the effect of finite gates we implement a real-space
Poisson solver with multigrid techniques\cite{mult1}.

\begin{figure}
\includegraphics[width=7.0cm]{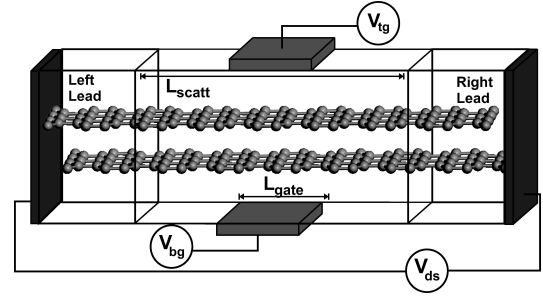}
\caption{\label{fig1} Schematic view of a dual-gated bilayer
graphene.}
\end{figure}

In Fig. \ref{fig1} we show a schematic representation of the bilayer
dual-gated graphene nanotransistor considered in our calculations.
The atomic geometry is composed of an $AB$ stacked bilayer graphene
(BG) with a scattering region  up to $15$ nm in length, sandwiched
by left and right leads. A source-drain voltage ($V_{ds}$) is
applied across the system defining the current direction. The charge
transport properties were calculated with the NEGF-DFT formalism,
which uses the Kohn-Sham Hamiltonian and overlap matrices computed
with the \textsc{siesta} code\cite{siesta} to self-consistently obtain the
density matrix. In order to investigate a finite dual-gate system,
we have done two main modifications in the \textsc{transampa} code
\cite{transampa}: (i) The inclusion of $k$ points that are
transversal to the transport direction ($k_{\perp}$), where for each
$k_{\perp}$ point in reciprocal space we find a
$k_{\perp}$-dependent density $D(k_{\perp})_{\mu\nu}$, by means of
an integration of the lesser Green's function\cite{dft_negf,ref39},
and transmittance $T(E,k_{\perp})$, given by:\cite{negf4}
\begin{equation}
T(E,k_{\perp})=Tr\left[\Gamma_{L}^{k_{\perp}}(E)G^{a}(E,k_{\perp})\Gamma_{R}^{k_{\perp}}(E)G^{r}(E,k_{\perp})\right].
\end{equation}
$G^{r(a)}(E,k_{\perp})$ is the retarded (advanced) Green's function,
whereas, $\Gamma_{L(R)}^{k_{\perp}}(E)$ are the terms that couple
the leads to the scattering region. The total density and
transmittance are obtained upon an integration in the reciprocal
space with the following equations:
\begin{equation}
D_{\mu\nu}=\frac{1}{(2\pi)^{2}}\int dk_{\perp}^{2}D(k_{\perp})_{\mu\nu},
\end{equation}
\begin{equation}\label{transmit}
T(E)=\frac{1}{(2\pi)^{2}}\int dk_{\perp}^{2}T(E,k_{\perp}).
\end{equation}
These modifications allow the simulations of truly 2D systems. Once
the density matrix convergence is achieved, the current is
calculated within the Landauer-B\"uttiker\cite{negf1} model in the
non-interacting approach of Meir-Wingreen\cite{negf4}, where the
current is given by:
\begin{equation}\label{current}
I=\frac{2e}{h}\int_{-\infty}^{+\infty}T(E)\left[f(E-\mu_{L})-f(E-\mu_{R})\right]dE,
\end{equation}
$T(E)$ is given by (\ref{transmit}), and
$f(E-\mu_{L/R})$ are the Fermi-Dirac distributions for the
left(L)/right(R) leads. We used the local density approximation
(LDA) for the exchange-correlation functional\cite{pz} since it
correctly describes the graphene interlayer distance without the
inclusion of van der Waals corrections.\cite{vdw} (ii) The second
modification was the inclusion of a real-space Poisson solver that
allows non-periodic solutions for the Hartree potential, necessary
to set different values for the top ($V_{tg}$) and back ($V_{bg}$)
gate voltages with finite gate lengths. Therefore, we solve the
Poisson equation in a rectangular box, fixing the upper and lower
boundary conditions, where the values of $V_{tg}$ and $V_{bg}$ are
defined only in a finite region of length $L_{gate}$ (see
Fig. \ref{fig1}). In our investigation we consider $L_{gate}=1, 2, 3,
4, 5$, and $10~nm$, $L_{scatt}\approx 15~nm$, and a distance between
the upper and lower boundaries of $20$~\AA.

We guarantee the convergence of the transmittance using a
double-$\zeta$ basis, $300~Ry$ for the mesh cut-off and 800
$k_\perp$ points in the Brillouin zone with the Monkhorst-Pack
scheme\cite{monkhost}. We used a fully relaxed geometry with a
carbon-carbon distance of $1.428$~\AA~ and an interlayer distance of
$3.203$~\AA. In the transport calculations, a tiny imaginary value
of $10^{-6}~Ry$ is added to the energy in order to calculate the
retarded Green's function, and the electron density was calculated
with 60 energy points and 5 poles to evaluate the integral of the
lesser Green's function.

\begin{figure}
\includegraphics[width=8.5cm]{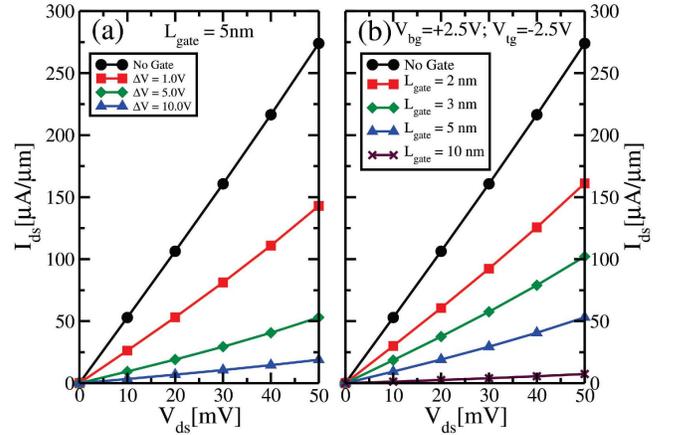}
\caption{\label{fig2} (Color online) In (a) we present $I_{ds}\times V_{ds}$ curves varying 
                      $\Delta V$, for $300K$ and $L_{gate}=5.0nm$. In (b) 
                      $I_{ds}\times V_{ds}$ curves varying $L_{gate}$, for $300K$ and $\Delta V=5.0V$.}
\end{figure}

The flowing current $I_{ds}$ as a function of the source-drain
voltage $V_{ds}$ is depicted in Figs. \ref{fig2}(a) and 2(b) for room
temperature. In our simulations the gate voltages $V_{bg}$ and
$V_{tg}$ are tuned independently. In Fig. \ref{fig2} all curves are
for gate voltages with $V_{bg}=-V_{tg}$, where the voltage
difference is $\Delta V = 2|V_{bg}|$. In Fig. \ref{fig2}(a) we used
$\Delta V=1.0$, $5.0$ and $10.0~V$, generating electric fields of
$0.5$, $2.5$, and $5.0~V/nm$, comparable to experimental values. In
all experiments\cite{on-off,tunable2,prl_105_166601}, the electric
displacement field in the bilayer is estimated through the
dielectric constant and thickness of the substrate dielectric layer,
and from the effective potential caused by the carrier doping due to
the experimental conditions. In this Brief Report, the electric field was
calculated via $\Delta V/d_{bt}$, where $d_{bt}$ is the distance
between the bottom and top gates. For the configuration studied 
here, we verify a decrease of the current as $\Delta V$
increases, suggesting the opening of a band gap that increases with
$\Delta V$. Moreover, for fixed gate length and gate voltages, the
electric current has a linear dependence with the source-drain
voltage, indicating an ohmic contact between the gated and non-gated
bilayer graphene. This behavior is also observed
experimentally\cite{on-off,tunable1,tunable2,tunable3}. From
Fig. \ref{fig2}(a) we show that we do not reach a zero electric
current, even for $\Delta V=10.0~V$. This result is in agreement with
experimental works that fabricate dual-gate bilayer graphene
devices\cite{on-off}.

In Fig. \ref{fig2}(b) we fixed the $V_{bg}=+2.5~V$ and $V_{tg}=-2.5~V$,
and we present the dependence of the flowing current with the gate
length $L_{gate}$, for $L_{gate}=2~nm$, up to $10~nm$. We
also observe a decrease of the current as $L_{gate}$ increases,
indicating an increase of the region with a gap. Again, the system
does not reach a zero current even for $L_{gate}=10nm$.

\begin{figure}[h]
\includegraphics[width=8.5cm]{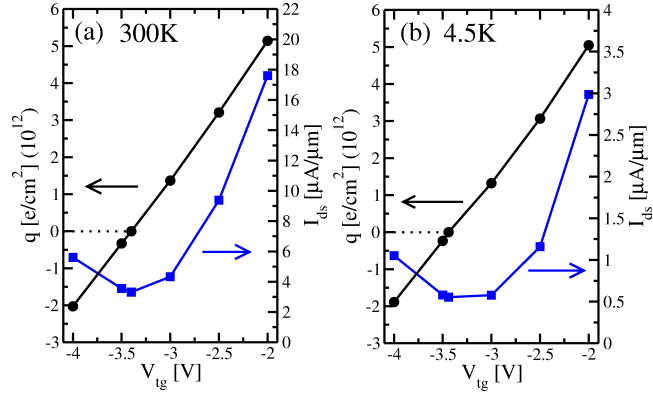}
\caption{(Color online) Electronic transport characteristic of a 
         bilayer graphene dual-gate FET, for $L_{gate}=5nm$ and $V_{bg}=+2.5V$. 
         The current (blue squares) and net charge (black circles) 
         at (a) room temperature and at (b) $T=4.5K$.\label{fig3}}
\end{figure}

Since we used open-boundary conditions, with the chemical potential
fixed by the leads, we can independently vary $V_{bg}$ and $V_{tg}$
to control the carriers at the scattering region. The presence of
the external electric field causes a breakdown of the equivalence
between the two layers. As a result, there is the opening of a band
gap and an inter-layer charge transfer. If there were no charge
transfer, the condition $V_{bg}=-V_{tg}$ would define the charge
neutrality point (CNP). However, the presence of a Hubbard-like
correlation term\cite{tb-matheus}, even at the mean-field level,
coupled to this charge transfer, leads to an overall shift of the
bands. As a result, the CNP occurs for $V_{bg}\neq-V_{tg}$, as shown
in Fig. \ref{fig3}.

Thus, we performed calculations where we fixed the $V_{bg}$ at a
certain value (+2.5 V) and varied the $V_{tg}$. The calculated
source-drain current and net carrier concentration in the bilayer,
as a function of the top gate, is shown in Figs. \ref{fig3}(a) and
\ref{fig3}(b) for a gate length $L_{gate}=5~nm$, for $T=300$ and
$4.5~K$, respectively. The bias voltage was fixed at $V_{ds}=10~mV$.
The current minimum in both curves corresponds to the charge
neutrality point. For a top gate voltage higher than $\approx -3.4~V$,
electrons are being injected in the scattering region, whereas for
$V_{tg}$ below this value the holes are being injected. As we have no
external doping, the chemical potential at the gated graphene is at
the midgap. Moreover, since the conduction and valence bands are very
similar, temperature will create intrinsic carriers but it will not
change the position of the chemical potential. Thus, there will be
some charge transfer from the graphene leads (electrons or holes)
depending on the shift of the mid-gap position, at the gated
graphene region, relative to the chemical potential of the leads.
Also, in Figs. \ref{fig3}(a) and \ref{fig3}(b) we plot the carrier concentration per
unit area. This presents a linear behavior as a function of the
variation of the gate voltage, with zero net charge at the charge
neutrality point. This indicates a linear displacement of the bands
with the gate voltage. There is a notable decrease in the current when
the temperature is diminished from $300$ to $4.5~K$. This is caused
both by variations in the transmittance with temperature as well as
by changes in the tail of the Fermi-Dirac distribution [see Eq. 
(\ref{current})]. For the case presented in Fig. \ref{fig3}, the
current decreases by more than six times. However, the net charge
concentration is very similar in both temperatures, a result of the
invariance of the mid-gap position with temperature, as discussed
above.

\begin{figure}
\includegraphics[width=7.0cm]{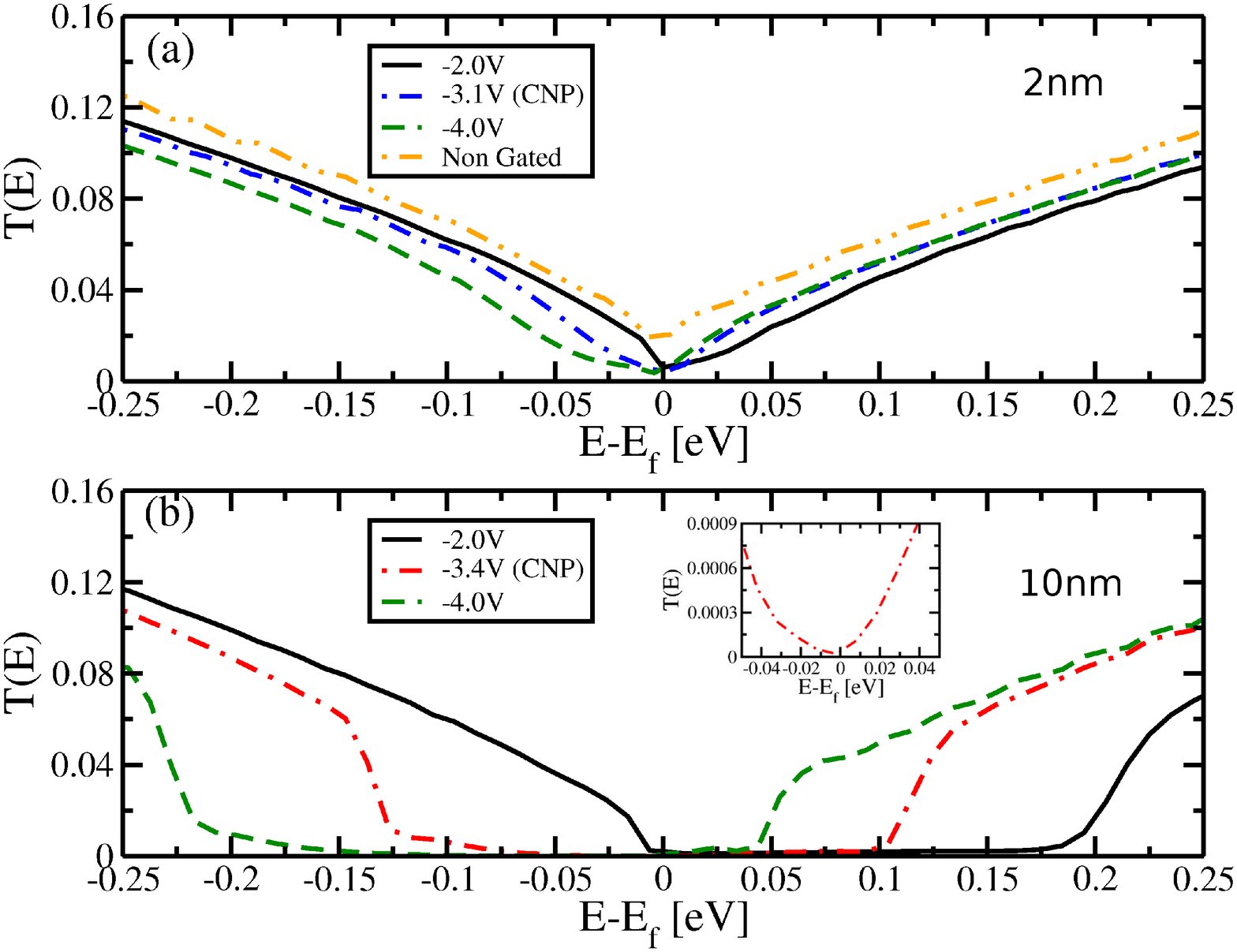}
\caption{\label{fig4} (Color online) (a) Transmittance for $L_{gate}=2~nm$ with $V_{tg}=-2.0~V$(solid black line),
                          $V_{tg}=-3.1~V$ (blue dashed-dotted line), $V_{tg}=-4.0~V$ (dashed green line), and the
                          nongated case (orange dashed-double-dotted line).
                      (b) Transmittance for $L_{gate}=10~nm$ with $V_{tg}=-2.0~V$ (solid black line),
                          $V_{tg}=-3.4~V$ (red dashed-dotted line) and $V_{tg}=-4.0~V$(dashed green line). 
                          Inset: Detail of the transmittance for $L_{gate}=10~nm$ at CNP close to the
                          Fermi energy region.}
\end{figure}

Figure \ref{fig4} shows the effect of the gate potential in the transmittance 
for (a) $L_{gate}=2~nm$ and (b) $L_{gate}=10~nm$; at $300~K$ with a fixed 
back gate voltage of $+2.5~V$, and varying $V_{tg}$. Under the same conditions, 
for larger gates the modifications on the transmittance are more expressive
in comparison to the smaller ones.
For each $V_{bg}/V_{tg}$ configuration the gate potentials causes a low transmittance region that is much
more pronounced with $L_{gate}=10~nm$ in comparisson with $L_{gate}=2~nm$.
However, the transmittance never goes to zero, as can be seen on the inset in Fig. \ref{fig4}(b).

\begin{figure}
  \includegraphics[width=7.0cm]{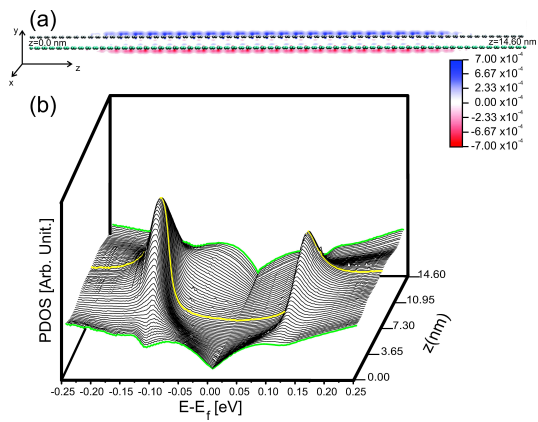}
  \caption{\label{fig5} (Color online) (a) Difference between the non-gated and gated local charge density. 
                        As can be seen, there is a transfer of electrons from the top to the botton layer (units of $e/bohr^{3}$).
                        (b) Projected density of states every four atoms (equivalent to the unitary cell) along the $z$ direction. 
                        The system considered had a $L_{gate}$ of $10nm$ and was at the CNP.}
\end{figure}

It is important to emphasize that in our calculations the electric
field is applied only in a finite portion of the scattering region
through the dual-gate system. As a consequence, the inter-layer
charge transfer caused by this electric field occurs only in the
gated region. In Fig. \ref{fig5}(a) we present the charge induced
at CNP, by a dual gate with $L_{gate}=10~nm$. In the bottom (top)
layer there is an excess (lack) of electrons. In Fig. \ref{fig5}(b) we 
show the density of states projected on every four atoms (equivalent to the unitary cell) 
along its entire length, from the left-hand ($z=0.0~nm$) to the right-hand electrodes ($z=14.6~nm$).
Near to the leads (green lines) the projected density of states (PDOS) 
are similar to the PDOS of a non-gated pristine bilayer, as expected, 
whereas directly under the gated region (yellow line) the PDOS is characteristic 
of a bilayer with an applied perpendicular electric field, where the two peaks 
located at approximately $\pm 12~meV$ are associated with the presence of a "mexican" hat behavior of the bands. 
However, since the gate is spatially finite, even in the presence of an energy band gap 
there is a finite density of states present. This is due to the penetration of the 
wave function in this forbidden energy region. This results explains the absence of an energy gap for transport [{\it e.g.}. see Fig. \ref{fig4}(b).]

\begin{figure}
\includegraphics[width=8.5cm]{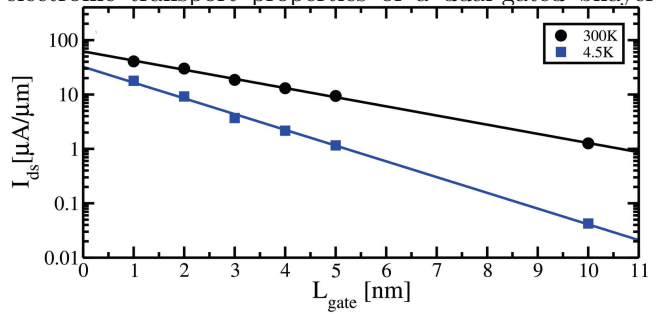}
\caption{\label{fig6} (Color online) $I_{ds}\times L_{gate}$ curves 
            for $\Delta V=5.0V$ and $V_{ds}=10mV$, for $300K$ and $4.5K$. 
            For all graphs, $\Delta V=2|V_{tg}|$ and $V_{bg}=-V_{tg}$.}
\end{figure}

To understand further the persistent current on the device, we look at its behavior as a 
function of the gate length. In Fig. \ref{fig6}, we fix $\Delta V = 5~V$ and $V_{ds} = 10~mV$, and vary $L_{gate}$ for $300$
and $4.5~K$. We obtain an exponential dependence between the current and the gate length, characteristic of a tunneling regime. 
The tunneling current persists even for the largest gate length, $L_{gate} = 10nm$. A significant increase 
of the current with temperature can be observed; this is largely caused by the broadening of the transport window via 
the $[f(E-\mu_{L})-f(E-\mu_{R})]$ term. At the low-bias regime, the tail of the Fermi-Dirac 
function mostly controls the current, since there is a small density of states 
(and hence a small number of transport channels) around the Fermi level in the electrodes. 
It is important to remind that in our {\it ab initio} calculations we retain all the intrinsic 
properties of the device, without the presence of defects, impurity carrier doping, or a substrate
dielectric material. Therefore, the off-current in our simulations is explained by tunneling across the 
gated region, and strongly depends on the transport energy window caused by the Fermi-Dirac distribution.

For our nanodevice we estimated the on/off current ratio,
$(I_{on}/I_{off})$ at room temperature, $(300~K)$, and low
temperature, $(4.5~K)$. The {\it on} current is calculated for the
non-gated bilayer graphene with a source-drain voltage of $10~mV$ and
the {\it off} currents are all defined at the charge neutrality
point, with a back gate voltage of $+2.5~V$ and the source-drain voltage
fixed at $10~mV$. The largest $I_{on}/I_{off}$ happens at
$L_{gate}=10~nm$, which is around $100$ at room temperature, and
$1250$ at $4.5~K$.

In summary, using {\it ab initio} calculations, we study the electronic transport properties 
of a dual-gated bilayer graphene nanotransistor. Under the action of an external electric field, we show that, 
even for gate lengths up to  $100$~\AA~ and for room temperature and $4.5~K$, a non-zero current is exhibited. 
The results can be explained by the presence of a tunneling regime due the remanescent states in the gated region.

We would like to thank E. Mucciolo for comments and
discussions. This research was supported by Brazilian agencies
INCT/CNPq and FAPESP.

\end{document}